\documentclass[12pt]{article}
\usepackage{setspace}
\usepackage{graphicx}
\usepackage{caption}
\usepackage{subcaption}
\usepackage{fullpage}
\usepackage[utf8]{inputenc}
\usepackage[T1]{fontenc}
\usepackage{amssymb,amsmath}
\usepackage{natbib}
\usepackage{soul}
\usepackage[hyphens]{url}
\usepackage[pagebackref=FALSE,pdfstartview=FitH,colorlinks,citecolor=black,filecolor=black,linkcolor=black,urlcolor=black]{hyperref}
\usepackage{cleveref}
\usepackage{xcolor}
\usepackage{booktabs}
\usepackage{gensymb}
\usepackage{pgfplots}
\pgfplotsset{compat=newest}
\usepackage{amsmath}
\usepackage{xcolor}
\pgfplotsset{every axis/.append style={
    label style={font=\small},
    tick label style={font=\scriptsize}
}}

\newcommand{\decomposeSundayAfternoon}{8\%}
\newcommand{\decomposeGrocery}{10\%}
\newcommand{\decomposeCrossBorder}{6\%}
\newcommand{\decomposeIncremental}{58\%}
\newcommand{\decomposeUnexplained}{17\%}

\DeclareRobustCommand{\rev}[1]{{#1}}

\title{

Opening Hours and Consumer Behavior:\\
Evidence from GPS Data and Deregulation\thanks{
We are grateful to Lucca Horta for excellent research assistance.
We thank seminar participants at KU Leuven, the UK Network of Industrial Economists annual conference and the 4th Workshop on UK Digital Economics for helpful comments and suggestions.
This research was partially funded by the British Academy/Leverhulme Research Grant.
}
}
\author{
Javier D. Donna\footnote{University of Miami, Herbert Business School and Rimini Center
for Economic Analysis, \url{jdonna@miami.edu}}
\and
Marit Hinnosaar\footnote{University of Nottingham and CEPR, \url{marit.hinnosaar@gmail.com}}
\and
Toomas Hinnosaar\footnote{University of Nottingham and CEPR, \url{toomas@hinnosaar.net}}
\and
André Trindade\footnote{Nova School of Business and Economics, \url{andre.trindade@novasbe.pt}}
}
\date{June 21, 2026}

\begin{document}
\def\pretrendPsunmorning{0.528}
\def\pretrendPsunaftern{0.330}
\def\pretrendPborder{0.000}
\def\pretrendPfood{0.051}

\maketitle

\begin{abstract}
In 2019, North Dakota repealed its Sunday closing law, which had required most non-grocery stores to close between midnight and noon. Using this policy change and consumer GPS data, we study the impact of opening hours on shopping behavior and welfare. We compare visits before and after the repeal in North Dakota and neighboring states using difference-in-differences and event-study designs. The repeal caused a large increase in Sunday morning visits, originating partly from intertemporal, store-type, and cross-border substitution. The closing law's welfare loss is equivalent to increasing the travel distance to affected stores by about 1.4 miles per consumer.
\end{abstract}



\section{Introduction}

Store opening hours are regulated in many countries to either protect workers, promote fair competition between large and small businesses, or preserve traditional weekend activities such as church attendance and family time. However, such restrictions also constrain consumer behavior. When stores are closed, consumers have to adjust by shopping at less convenient times or at less convenient stores, traveling farther, or deferring purchases altogether.
\rev{
The economic importance of store hours is not limited to formal regulation. Retail hours also change through firms' own operating decisions. The proliferation of 24-hour stores over the past half century, most notably Walmart in the United States, fundamentally altered when consumers could shop \citep{biddle_twenty-four_2026}.
Despite the large changes in store opening hours and political salience of the related regulation, relatively little is known about the impact on consumers.
}

In this paper, we study how store opening hours affect consumer shopping behavior and welfare. A major empirical challenge in studying this question is that firms optimally choose opening hours in response to their anticipated consumer demand. This creates a simultaneity problem: stores are open when consumers want to shop, and consumers shop when stores are open. To understand the causal impact of opening hours, one needs a setting where store opening hours vary exogenously relative to consumer preferences.
To overcome this difficulty, we exploit a natural experiment: the repeal of North Dakota’s Sunday closing law on August 1, 2019. Before the repeal, the law required most non-grocery stores to close between midnight and noon on Sundays, making it one of the strictest blue laws in the United States. The repeal removed this constraint, allowing stores, notably general merchandise stores such as Walmart, to open earlier on Sundays.

Specifically, in this paper we ask two main questions: \textit{First}, what is the impact of store opening hour restrictions on consumer shopping behavior, i.e., do consumers respond by changing where and when they shop? \textit{Second}, what are the welfare implications to consumers of the restriction?

To answer these questions, we use detailed Global Positioning System (GPS) data from mobile devices to track consumer visits to retail locations at an hourly frequency and at the store level. The data, provided by Advan Research, are based on anonymized location information from a large panel of mobile phone applications.\footnote{The mobility data were previously provided by SafeGraph and transitioned to Advan Research in 2023. The SafeGraph data were widely used to study movements during COVID-19. The representativeness of the SafeGraph data has been evaluated by \cite{chen_geographic_2020} and \cite{athey_estimating_2021}.}
The dataset records the number of visits to each store by hour of the week, allowing us to analyze Sunday morning shopping and substitution patterns. We also observe monthly visits to each store by the home census block group of visitors, enabling us to study travel distances. Our sample includes visits to stores in North Dakota and its neighboring states before and after the deregulation in 2019.

We focus our analysis on the largest general merchandise store chain, Walmart, and grocery stores. We do so for two reasons: first, while we do observe store visits before and after the policy change, we do not directly observe whether each store responded to deregulation by increasing the number of hours it was open. In the case of Walmart, we know that all stores in North Dakota were closed on Sunday mornings before the deregulation and \rev{that most stores were open on Sunday mornings immediately} after the deregulation. Second, Walmart operates many stores that are virtually identical across states, allowing us to use other states as a control group.
We analyze visits to grocery stores because these were not directly affected by the regulation, as they were already allowed to be open on Sunday mornings before the policy change.

We first document the causal effect of deregulating Sunday morning store hours on consumer shopping behavior. Using difference-in-differences and event-study designs, we compare hourly store visit patterns before and after the repeal in North Dakota with those in neighboring states that did not experience a regulatory change. We find that the repeal led to a substantial increase in Sunday morning visits to Walmart stores in North Dakota.
While the direction of this effect is intuitive, our contribution lies in quantifying its magnitude and decomposing consumer responses across distinct margins of adjustment.
This decomposition matters for policy evaluation: restrictions that merely shift shopping across time impose different welfare costs than those that shift spending across jurisdictions or store formats. Understanding these magnitudes is essential for the debates over Sunday trading laws that remain active in Europe and, \rev{as the 2025 legislative effort in North Dakota (a proposal to reinstate similar restrictions, which was ultimately defeated) demonstrates,} in the United States.

Second, we examine how consumers adapt when shopping opportunities are restricted. If an individual's favorite store is not open at the time when they would prefer to shop, what happens? Will they change to a new store? Shop at a different time? We find that the added flexibility to shop on Sunday mornings reshaped consumer behavior in three ways: across time, across space, and across store type. First, visits slightly declined on Sunday afternoons. Second, border Walmart stores in Minnesota saw fewer Sunday-morning visits, as North Dakota residents no longer had a reason to cross the state line. Third, grocery stores, which were exempt from the restrictions, experienced a drop in visits after the repeal.
Attributing the overall increase in Walmart visits following deregulation, we find that \decomposeGrocery{} came from grocery stores that had remained open under the old law, \decomposeSundayAfternoon{} came from temporal substitution from Sunday afternoons, and \decomposeCrossBorder{} came from consumers who had previously crossed into Minnesota. The remaining share reflects a combination of additional visits, substitution from other times of the week, and substitution from stores outside our sample.

To quantify the welfare implications of increased flexibility in shopping hours, we present a discrete choice model where consumers choose their preferred store and shopping time. The model captures preferences over shopping time, store type, and travel distance. To overcome the lack of pricing information across stores, we focus on travel distance as our main welfare metric. We find that the Sunday morning ban was welfare-equivalent to Walmart stores being about 1.4 miles further away for each consumer.

We make three novel contributions. First, we document the causal effect on consumer shopping behavior of repealing the blue law prohibiting Sunday morning sales. In doing that, we provide evidence of intertemporal and spatial substitution. Second, we quantify the value to consumers from the additional flexibility of choosing when to shop after the deregulation. We find that the consumer welfare loss due to the Sunday prohibition is equivalent to increasing the distance to the studied stores by 1.4 miles. Third, we use detailed geolocation data that has not been used before to analyze the impact of changes in regulation in the retail market.

\paragraph{Related literature.}
Our paper is related to several strands of literature. First, we contribute to the literature on regulation and, in particular, store-hour regulation. Opening hours restrictions can have both social benefits and costs. Earlier empirical work has focused mainly on potential benefits, examining effects on employment \citep{skuterud_impact_2005, paul_after_2015, bensnes_earning_2019, rizzica_effects_2023} and church attendance \citep{gruber_church_2008, cohen-zada_religious_2011}.
These studies show that opening hours restrictions can protect workers and even strengthen community and religious practices.  Less is known about the costs to consumers, such as the inconvenience from restricted store access. Measuring these costs is difficult because it requires data on shopping time and location at an hourly level. To the best of our knowledge, our paper is the first to use GPS data to study the impact of store-hour deregulation. \cite{jacobsen_timing_2005} use time diary data in the Netherlands to document that shopping activities shifted toward workday evenings after deregulation.
Our GPS dataset allows us to study both shopping time and location choices. Moreover, as the policy change affected only certain types of stores in one state, we have natural control groups (neighboring states and other types of stores) that allow us to study the causal impact of the deregulation.

The literature has also looked at the impact of blue laws on alcohol consumption \citep{bernheim_consumers_2016, hinnosaar_time_2016}. A related body of work examines how extending retail hours affects the consumption of specific products. Studies of alcohol sales restrictions consistently find that extended hours increase alcohol consumption and associated harms (see \cite{popova_hours_2009} for a review). Our setting differs in that the Sunday restriction applied to general merchandise rather than a product with addiction or health externalities, but these findings underscore that hour restrictions do bind consumer behavior.

Our paper is also related to the literature using mobile phone location data to study consumer behavior.\footnote{The literature using mobile phone location data has also studied other topics such as policing \citep{chen_smartphone_2023}, voting \citep{chen_racial_2022}, political protests \citep{sobolev_news_2020}, geographic mobility \citep{chen_geographic_2020}, and mobility and economic activity \citep{kreindler_measuring_2023}.} The literature has analyzed preferences for restaurants \citep{athey_estimating_2018, charles_consumer_2025} and cars \citep{yavorsky_consumer_2021}, racial segregation in consumption \citep{athey_estimating_2021}, and consumption externalities from remote work \citep{miyauchi_economics_2021}. To the best of our knowledge, we are the first to use mobile phone location data to analyze the impact of retail deregulation.

We also contribute to the literature on store choice. The literature has documented how consumers choose stores based on prices \citep{bell_shopping_1998, thomassen_multi-category_2017, paciello_price_2019}, convenience and distance \citep{taylor_food_2016, marshall_measuring_2018, eizenberg_retail_2021, huang_consumer_2023}, product quality \citep{matsa_competition_2011}, and other non-price characteristics \citep{smith_supermarket_2004, smith_store_2006, trindade_price_2015, ellickson_measuring_2020}. We show that store opening hours are another important factor in store choice.\footnote{The effect of temporary store closures has been studied by \cite{kotschedoff_persistence_2025}.}

Our paper is also related to the literature that focuses on the impact of discount stores on the grocery market. The literature has documented that upon entry, large discount stores take the market share from grocery stores.\footnote{This is opposed to the documented positive externalities of department stores in malls \citep{gould_contracts_2005}, small stores in Dutch shopping streets \citep{koster_shopping_2019}, and car dealerships \citep{murry_consumer_2020}. Heterogeneous effects of store externalities have been shown by \cite{vitorino_empirical_2012}.} This has been shown in the case of Walmart \citep{basker_causes_2007, hausman_consumer_2007, ellickson_wal-mart_2013} and dollar stores \citep{caoui_impact_2022} in the US, and discount stores in other countries \citep{evensen_co-location_2024}.\footnote{Relatedly, \cite{jia_what_2008} showed that in the 1980s and 1990s, the expansion of Walmart led to the exit of a large percentage of small discount stores. Similarly, \cite{talamas_marcos_surviving_2025} showed that the entry of chain convenience stores had negative effects on independent convenience stores.} We document a similar substitution effect toward Walmart, not through the entry of a new store, but through the extension of opening hours.

Our work is related to the literature quantifying the extent of cross-border shopping. The literature has mainly focused on cross-border shopping motivated by differences in prices (e.g., \cite{campbell_real_2004,  friberg_hump-shaped_2022}) or the regulations prohibiting the sale of certain products, like marijuana \citep{hansen_federalism_2020}.\footnote{Health and public economics literature has focused on cross-border shopping of alcohol and cigarettes driven by differences in taxes \citep{asplund_demand_2007, harding_heterogeneous_2012, johansson_cross-border_2014, hindriks_heterogeneity_2019}.} Our work shows that the sales restrictions on most non-food products during certain hours also lead to cross-border shopping.

Our work also relates to the studies investigating the opportunity cost of time in grocery shopping.\footnote{The literature on ride-hailing markets has studied intra-daily variation in the opportunity cost of time \citep{buchholz_personalized_2025}.}
The literature has shown that the opportunity cost of time changes over the business cycle \citep{nevo_elasticity_2019}, at retirement \citep{aguiar_consumption_2005}, and during unemployment \citep{bronnenberg_consumer_2024}.
We show that the opportunity cost of time also varies throughout a typical week.

\section{Institutional Background and Natural Experiment}

The Sunday closing laws, also known as blue laws, limit store opening hours or prohibit the sale of specific items on Sundays.
Sunday closing laws aim to either protect workers, promote fair competition between large and small businesses, or preserve traditional weekend activities such as church attendance and family time. The arguments against these laws are typically based on consumer welfare.

The store opening-hour regulations have deep historical roots. In A.D. 321, the Roman Emperor Constantine issued the first known prohibition on Sunday labor, declaring: ``On the venerable Day of the Sun let the magistrates and people residing in cities rest, and let all workshops be closed''.\footnote{Source: \url{https://en.wikipedia.org/wiki/Blue_law}, accessed Aug 27, 2025.}

Sunday closing laws remain widespread across European countries, including Germany, Austria, Switzerland, Norway, and the UK, among others,  though recent decades have seen substantial reforms. In England, the Sunday closing laws were relaxed in 1994. Until then, on Sundays, large shops had to be closed; since then, large shops have been permitted to open up to six hours. A 2015-16 proposal to let local authorities set their own rules for Sunday store opening hours was rejected in Parliament. Germany deregulated weekday store opening hours in 2006, but on Sundays still requires most stores to be closed. Several countries lifted the Sunday restrictions in recent years, including Denmark in 2012, and Finland and Hungary in 2016. On the other hand, Poland introduced one of Europe's strictest bans in 2018, closing almost all stores on Sundays. In Portugal, large retailers have been allowed to open on Sunday since 2010, but the matter is still under public debate (as recently as June 2025, a bill was introduced in the national parliament attempting to force large retailers to close again on Sunday).

In the U.S., most of the laws that place a wide ban on commerce on certain days have been repealed. However, many states still restrict the hours during which alcohol can be sold or ban the sale of cars on Sundays. In 2019, North Dakota repealed one of the country's strictest blue laws.

Before August 1, 2019, most stores in North Dakota were required to close from midnight until noon on Sundays. Exceptions included grocery stores (supermarkets and convenience stores), drugstores, flower shops, newsstands, and gas stations. The law also prohibited the sale of most non-food items on Sunday before noon, including clothing, appliances, kitchenware, and toys. Notably, department stores and discount stores, like Walmart, had to be closed.

North Dakota lawmakers and the governor who supported the repeal of the Sunday closing law argued that it would allow people to spend money locally rather than online or in stores in other states, thereby increasing tax revenue and employment.\footnote{See e.g. \url{https://www.usnews.com/news/best-states/articles/2019-03-26/north-dakotans-will-soon-be-able-to-shop-on-sunday-mornings} and  \url{https://apnews.com/article/5c255da5b6f14414a8d1fe291a77ce77}.} Anecdotal evidence suggests that until August 2019, many North Dakota residents did their shopping in Minnesota on Sundays. Shopping in Minnesota was viable for residents of two of the three largest North Dakota cities (Fargo and Grand Forks), which are located close to the border with Minnesota.

The Sunday closing law was a divisive issue in North Dakota. In the decade prior to the repeal, there were three unsuccessful attempts to repeal the law.
The main reasons the repeal finally passed include a generational shift among lawmakers; the election of a new governor; the addition of a clause preventing leases or franchise agreements from requiring Sunday openings; increased pressure from border-city retailers; and a growing realization that the ban did not protect store employees who were already stocking shelves but merely prohibited customer-facing work, not Sunday work itself.\footnote{See e.g.: \url{https://news.prairiepublic.org/local-news/2019-01-14/senate-majority-leader-reconsidering-opposition-to-blue-law-repeal}}
The 2019 repeal did not end the debate: in 2025, lawmakers considered reinstating similar restrictions, but the proposal was defeated.

The Sunday closing law repeal bill was signed in March 2019, and it took effect on August 1, 2019. Consumers were well aware of the change. In July, major retailers issued press releases listing their earlier Sunday opening hours. Local media reminded consumers that the first Sunday affected would be August 4, and published the new opening hours for the major retailers.

Retailers with nationwide opening-hour policies, such as Walmart, adjusted their North Dakota schedules to align with corporate policies. Before the repeal, North Dakota was the only state in which Walmart \rev{could} not operate 24 hours a day. Beginning in August 2019, Walmart stores in North Dakota started to be open \rev{on Sunday mornings}, just as they were elsewhere.\footnote{See e.g. \url{https://www.inforum.com/business/retailers-setting-earlier-sunday-openings-as-n-d-s-shopping-blue-law-ends-in-august}.} While large chains adjusted their hours, many independent stores did not. As we analyze how opening hours affect consumer behavior, we focus on Walmart to exploit the regulation-induced change in opening hours.

\section{Data}

\paragraph{Advan Research’s mobility data.}

To measure shopping behavior, we use store-visit data based on GPS geo-location information from mobile phones provided by Advan Research: Advan Monthly Patterns \citep{advan_monthly_2022} and Advan Weekly Patterns \citep{advan_weekly_2022}.\footnote{This data was made available by Advan Research (\url{https://advanresearch.com/}) on the Dewey Data platform (\url{https://www.deweydata.io/}).

}$^{,}$\footnote{According to Dewey, Advan's data is nearly identical in structure and quality to SafeGraph data, which was available in the past but transitioned to Advan in 2023, and which was widely used in prior research, especially to study movement patterns during COVID-19. The representativeness of SafeGraph data has been evaluated by \cite{chen_geographic_2020,athey_estimating_2021}.
}
The data are based on anonymized location information from a large panel of mobile phone applications. The applications track location with an accuracy of approximately 10 meters, and data collection occurs only with the explicit consent of the user.

The data consists of two separate aggregations of the mobile phone positioning information. First, it provides a store-level number of visits at a given hour of the week, for example, at 11 am on Sunday. This allows us to analyze temporal patterns of store visits. Second, it provides the store-level number of visits by consumer origin census block group. It allows us to analyze how shopping behavior depends on the distance to stores. Here, we do not observe the exact timing of the store visits. Instead, the data is aggregated across all hours.
For each store, we also observe its location, associated chain, and industry code, which allows us to separate grocery stores.

\begin{figure}[ht]
	\centering
	\includegraphics[width=\textwidth]{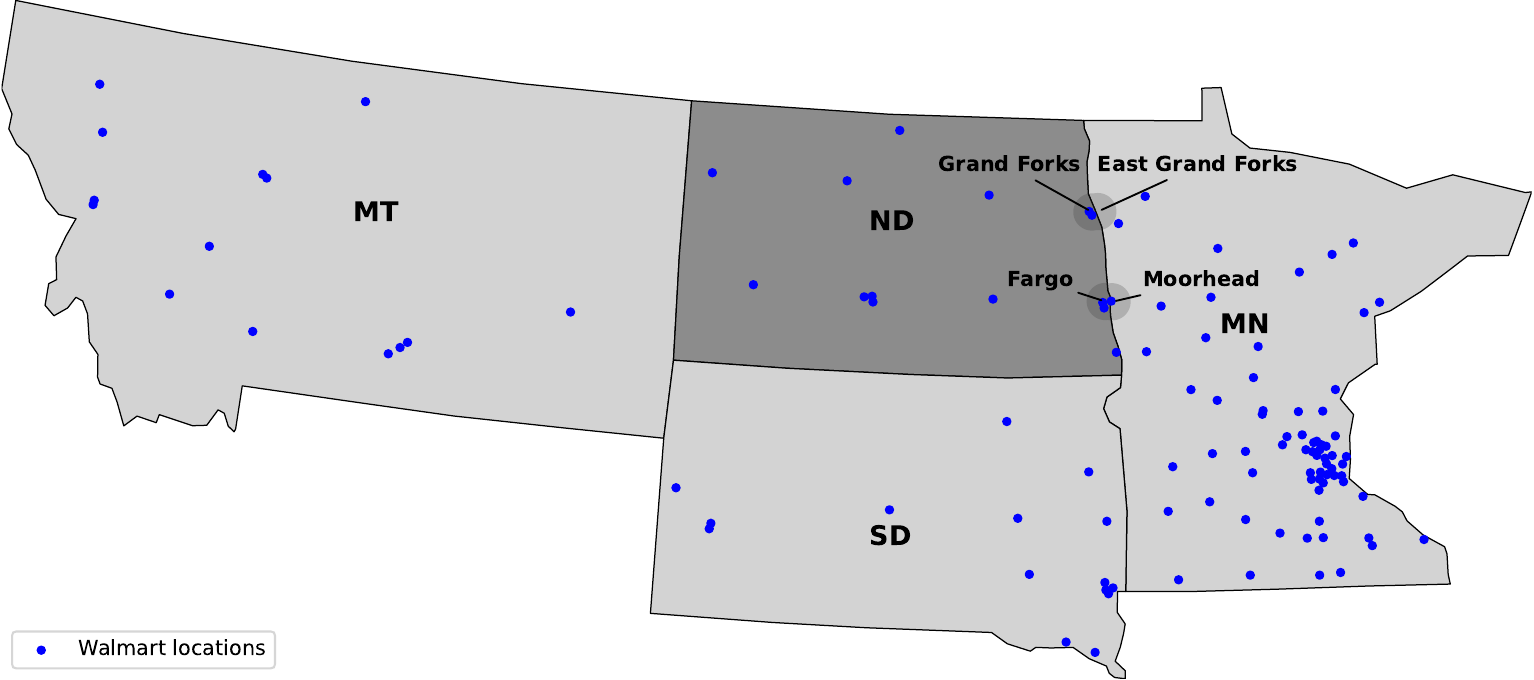}
	\caption{North Dakota and neighboring comparison states used in the analysis.
    Blue markers show Walmart superstore locations. Fargo-Moorhead and Grand Forks-East Grand Forks are twin-city pairs spanning the North Dakota-Minnesota border. For a full US map, see \Cref{fig:map_all_states} in \Cref{A:AdditionalTablesFigures}.
    }
	\label{fig:map_states}
\end{figure}

\paragraph{Sample and variables.}
We focus on North Dakota and its neighboring states---Minnesota, South Dakota, and Montana---which we use as a control group. These states are shown in Figure~\ref{fig:map_states}.
Our dataset includes visits to stores in 2019, before and after the deregulation.
We focus on the largest general merchandise store chain, Walmart, for two reasons. First, the retailer adjusted its opening hours in response to the policy change. Second, it operates many stores that are virtually identical across states. As a substitute for Walmart, we analyze visits to grocery stores (supermarkets and convenience stores). These were stores that were not directly affected by the regulation, as they were allowed to remain open on Sunday mornings even before the repeal of the law. However, grocery stores can be affected via spillovers. We restrict attention to grocery store chains with at least ten stores in these four states because, compared to independent stores, they were more likely to be open on Sunday mornings.\footnote{We also drop all stores identified in Advan Research's Shared Polygons data as sharing a polygon with another establishment: \rev{3 Walmart and 254 grocery stores}.}
\rev{In the main Walmart difference-in-differences sample, we exclude the two Minnesota Walmart stores located directly on the North Dakota border. These stores are analyzed separately in the cross-border substitution analysis because they were themselves affected by the North Dakota policy change through changes in cross-border shopping. Including them in the control group would therefore bias the estimated treatment effect upward.}

We define Sunday morning as the period from 6 am to noon. Although the policy also covered hours from midnight to 6 am, we exclude these because \rev{the nighttime consumer traffic is low,} and most detected visits during that time are likely to come from store employees rather than customers.\footnote{We obtain similar results when including these hours in the morning definition.} \rev{This is also consistent with the broader trend documented by \cite{biddle_twenty-four_2026}, who show that overnight retail hours have been in decline even outside regulatory contexts, suggesting that consumer demand during these hours is limited.
} We define Sunday afternoon as 1 pm to 2 pm. This period is the closest natural substitute for morning shopping trips while avoiding overlap with them. \rev{We start this window at 1 pm rather than noon because the noon to 1 pm hour is too close to the restricted period to provide a clean measure of substitution: before the repeal, consumers who had planned to shop in the morning could simply delay their trip until just after noon, and after the repeal, morning shoppers may still be inside the store shortly after noon. Starting the window at 1 pm provides a cleaner measure of substitution to a distinct afternoon shopping alternative.} In our data, the substitution effect is strongest in this early-afternoon window and diminishes rapidly thereafter, exactly as we would expect.

\begin{figure}[h!]
	\begin{center}
	\caption{The distribution of the percentage of visits on Sunday mornings, separately for stores in North Dakota and neighboring states before and after the change in regulation}

	\begin{subfigure}[b]{0.48\textwidth}
		\includegraphics[width=\textwidth]{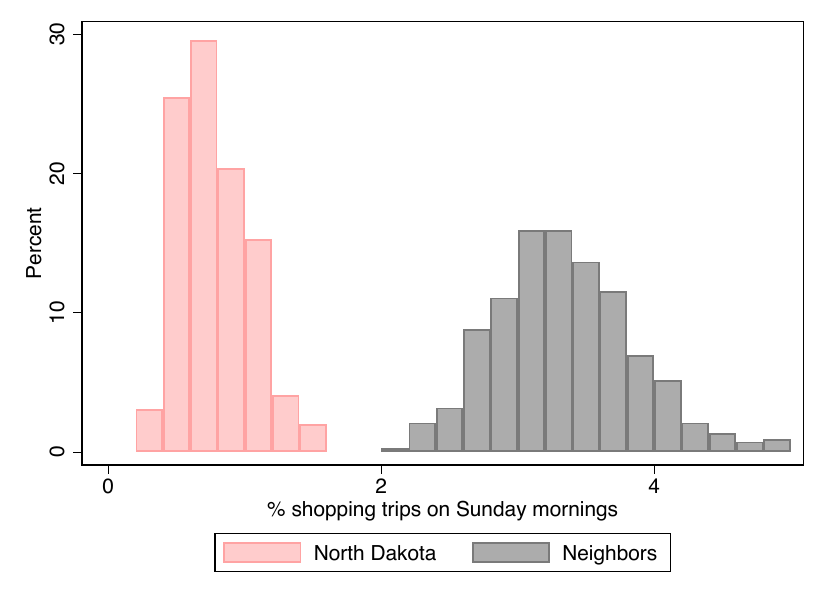}
		\caption{Walmart stores before policy change}
	    \label{gSunMornShareWalmart_JanJul}
	\end{subfigure}
	\hfill
	\begin{subfigure}[b]{0.48\textwidth}
		\includegraphics[width=\textwidth]{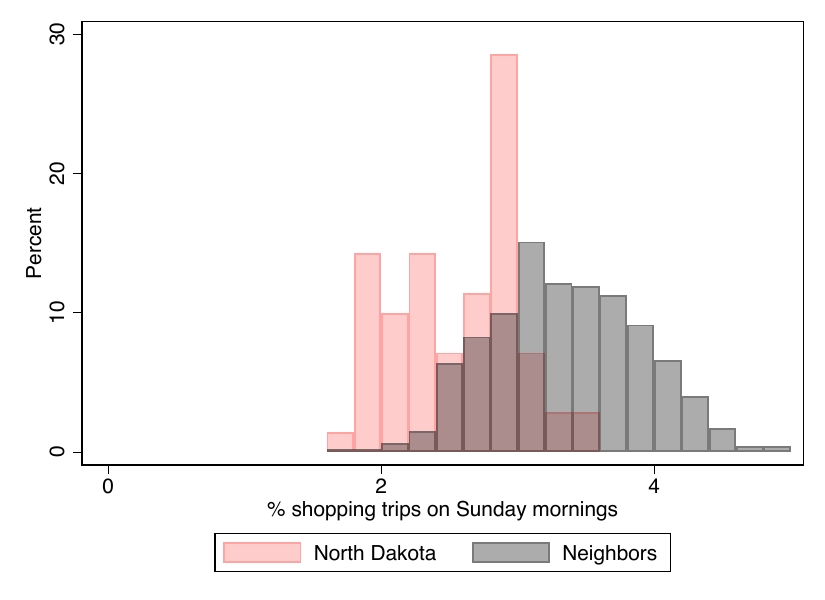}
		\caption{Walmart stores after policy change}
		\label{gSunMornShareWalmart_AugDec}
	\end{subfigure}

        \begin{subfigure}[b]{0.48\textwidth}
		\includegraphics[width=\textwidth]{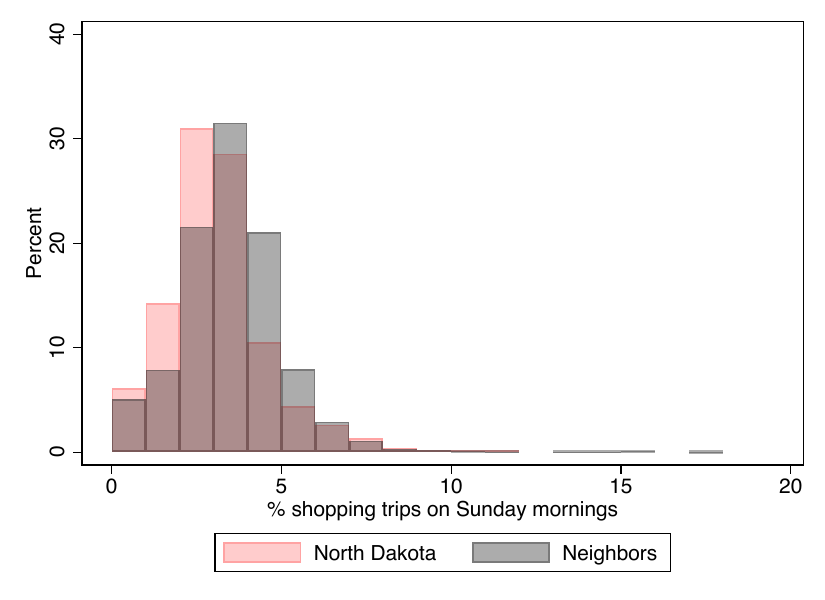}
		\caption{Grocery stores before policy change}
	    \label{gSunMornShareGrocery_JanJul}
	\end{subfigure}
	\hfill
	\begin{subfigure}[b]{0.48\textwidth}
		\includegraphics[width=\textwidth]{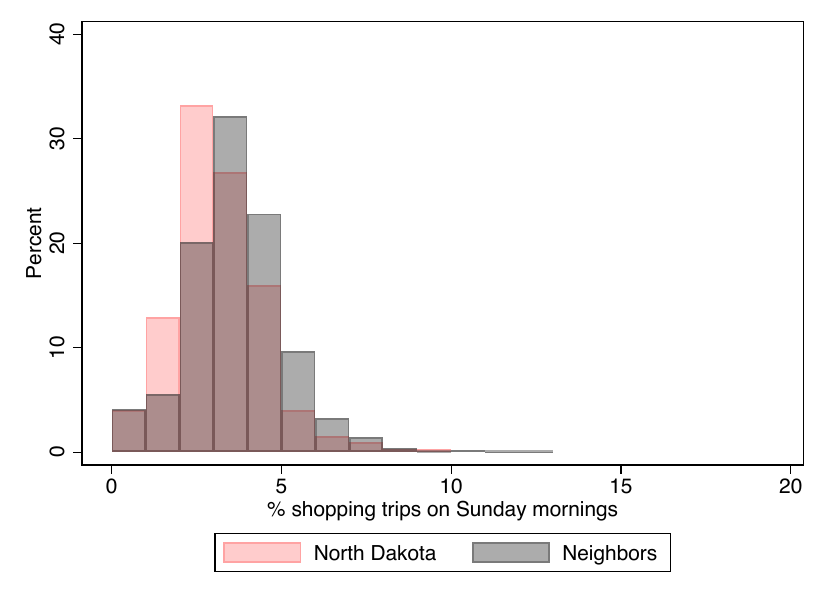}
		\caption{Grocery stores after policy change}
		\label{gSunMornShareGrocery_AugDec}
	\end{subfigure}

	\label{gSunMornShare}
	\end{center}
\footnotesize{
Notes: The figure presents the histogram of the percentage of shopping trips on Sunday mornings. An observation is a store-month pair.
}
\end{figure}

\paragraph{Summary statistics.}
The distribution of the percentage of store visits on Sunday mornings is in \Cref{gSunMornShare}. Before the reform, Walmart stores in North Dakota received a significantly smaller number of visits on Sunday mornings than stores in neighboring states (\Cref{gSunMornShareWalmart_JanJul}). Note that the average number of visits was close to, but not exactly, zero. This is because the data includes employees who were inside the stores even when the stores were closed to shoppers. After the reform, Walmart visits on Sunday morning increased, becoming more similar to those in neighboring states (\Cref{gSunMornShareWalmart_AugDec}).
To rule out that the change in shopping behavior does not correspond to a specific change in consumer preferences in August, \Cref{gSunMornShareGrocery_JanJul,gSunMornShareGrocery_AugDec} show that North Dakota consumers were shopping on Sunday mornings in grocery stores already before the reform, and the distribution is largely unaffected by the reform.
Overall, the figures indicate that the Sunday morning restriction was binding for consumers.
\Cref{tSummaryStatisticsReducedForm} reports summary statistics on store visits by store type and location prior to the reform.

\begin{table}[ht!]
\renewcommand{\tabcolsep}{3pt}
\begin{center}
\caption{Summary statistics}
\label{tSummaryStatisticsReducedForm}
\begin{footnotesize}
\begin{tabular}{ l cc }
\hline
\hline
 & North Dakota & Neighboring states\\ & (1) & (2) \\
 & \multicolumn{2}{c}{Panel A: Walmart stores} \\
Number of visits on Sunday mornings                         &     91.5&    368.7\\
Number of visits on Sunday afternoons                       &    199.7&    161.0\\
Number of visits per month                                  &   7856.1&   7374.1\\
Number of unique visitors per month                         &   3535.4&   3500.0\\
Number of stores                                            &       14&       92\\
 & \multicolumn{2}{c}{Panel B: Grocery stores} \\
Number of visits on Sunday mornings                         &     33.6&     46.2\\
Number of visits on Sunday afternoons                       &     12.3&     15.4\\
Number of visits per month                                  &    826.7&   1000.3\\
Number of unique visitors per month                         &    487.0&    579.6\\
Number of stores                                            &       65&      664\\
 \hline
 & At North Dakota border & Non-border\\ & (1) & (2) \\
 & \multicolumn{2}{c}{Panel C: Walmart stores in Minnesota} \\
Number of visits on Sunday mornings                         &    310.5&    380.6\\
Number of visits on Sunday afternoons                       &    112.9&    163.8\\
Number of visits per month                                  &   5235.9&   7578.9\\
Number of unique visitors per month                         &   2380.8&   3653.4\\
Number of stores                                            &        2&       63\\
\hline

\end{tabular}
\end{footnotesize}
\end{center}
\footnotesize{
Notes: The table reports average monthly statistics before the policy change (January to July 2019). In panel A, Walmart stores in neighboring states exclude two stores close to the North Dakota border. In panel B, grocery stores include both supermarkets and convenience stores.
}
\end{table}

\section{Empirical Analysis}
\subsection{Empirical Strategy}

\paragraph{Difference-in-differences.}
Our goal is to measure changes in consumer shopping behavior following the repeal of North Dakota's Sunday closing law. We estimate difference-in-differences regressions, comparing the change in store visits in North Dakota before and after the repeal to the corresponding change in neighboring states, where no such policy change occurred. We use difference-in-differences because a simple before-and-after comparison would ignore seasonal patterns that could bias the results. Our research design allows us to control for unobservable factors that are constant across all stores in a given month, or constant across all months in a given store.

Specifically, we estimate the following regression, where the outcome variable is the logarithm of the number of visits to store $i$ in month $t$ (or in specific hours or days):
\begin{align}
\label{E:DID}
	log Visits_{it}
	& = \beta \cdot NorthDakota_i \times AfterRepeal_t  \nonumber \\
	& + MonthFE_t + StoreFE_i + \varepsilon_{it}.
\end{align}

The causal interpretation requires that, absent the repeal, trends in Sunday morning visits would have evolved similarly in North Dakota and neighboring states. We assess this assumption by examining pre-treatment coefficients in our event-study specification and by conducting formal tests for differential pre-trends. As shown below, \Cref{fig:sunday_morn,fig:foodchain_sunday} show no systematic pre-trends.

The coefficient of interest is $\beta$, on the interaction term of the indicator whether the store is in North Dakota ($NorthDakota_i$) and the time period is after the repeal ($AfterRepeal_t$). It measures the change in the number of visits to the store $i$ in North Dakota after the repeal of the Sunday closing law, compared to the change in the number of visits in the neighboring states. The regression includes store and month fixed effects. We study the change in store visits on Sunday mornings, afternoons, and total visits. \rev{The samples used in the regressions are characterized in \Cref{tSummaryStatisticsReducedForm}. For the Walmart temporal regressions, the control group excludes the two Minnesota border stores as described above.}

\paragraph{Event study.}
To examine the dynamics of the policy effect and assess the validity of the parallel trends assumption, we complement the difference-in-differences analysis with an event study specification. This approach estimates the evolution of treatment effects over time by interacting period-specific indicators with the treatment group indicator. We replace the post-treatment indicator with a full set of time relative-to-repeal dummies, allowing us to trace the trajectory of store visits before and after the policy change. We estimate event studies separately by time block (e.g., Sunday mornings and Sunday afternoons) to visualize substitution patterns over time.

Three potential threats to identification merit discussion. First, if Walmart reallocated inventory or staffing from Minnesota border stores to North Dakota stores in response to the policy change, this could generate visit changes independent of consumer preferences. We view this as unlikely to drive our results because (a) the Minnesota border stores are small relative to Walmart's regional operations, and (b) we find no evidence of changes in total monthly visitors to these stores (\Cref{tRegrBorder_log}, column 3), only in the timing of visits. Second, anticipation effects could arise if retailers adjusted their behavior before August 2019. The event-study plots (Figures 3 and 4) show no evidence of gradual pre-trends consistent with anticipation; the change is sharp and coincides precisely with the repeal date. Third, as \cite{larcom_benefits_2017} emphasize, consumer responses to adding versus removing options may differ if consumers exhibit status quo bias. Our welfare estimates should therefore be interpreted as the value of the new option rather than the loss from hypothetically removing it.

\subsection{Shopping Time Effects}

We begin by analyzing how the extended store hours due to the repeal of the Sunday closing law affected the timing of store visits. Here, we focus on Walmart stores only, as all Walmart stores are homogeneous and were directly affected by the Sunday sales restrictions. \Cref{tRegrTimeSubs_walmart_log} reports the results from our difference-in-differences regressions (\Cref{E:DID}), where the outcome is the logarithm of the number of visits to a store in specific time blocks or the number of unique visitors in a month. For the former, we also present results controlling for the logarithm of the number of unique monthly visitors.
This controls for changes in the sample of mobile devices in the dataset. The underlying assumption is that while the deregulation might have affected when and how often a consumer visits a particular store, it did not change whether a consumer visits the store at least once a month.
All standard errors from the difference-in-differences and event-study specifications are clustered at the store level.

\begin{figure}[h!]
	\begin{center}
	\caption{Event study of the logarithm of the number of visits to Walmart stores in North Dakota versus neighboring states}

	\begin{subfigure}[b]{0.48\textwidth}
		\includegraphics[width=\textwidth]{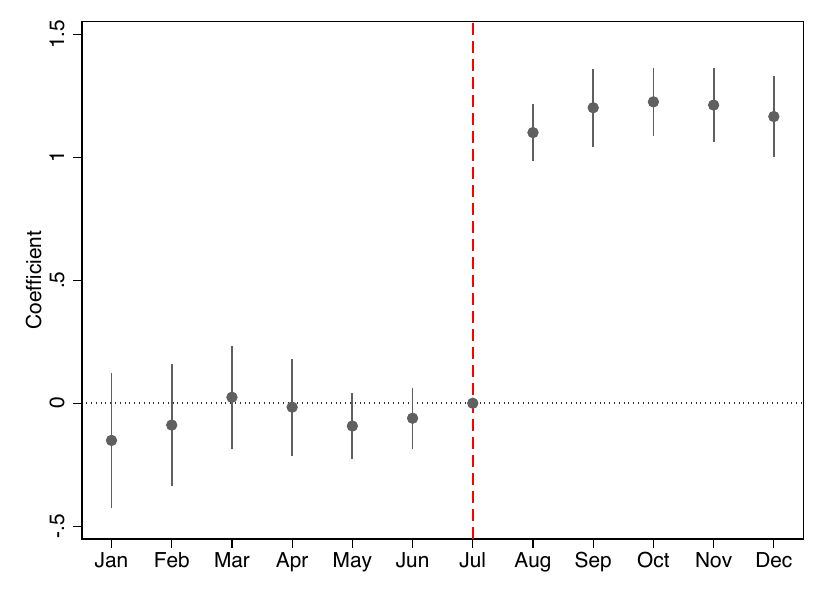}
		\caption{Sunday Morning}
	    \label{fig:sunday_morn}
	\end{subfigure}
	\hfill
	\begin{subfigure}[b]{0.48\textwidth}
		\includegraphics[width=\textwidth]{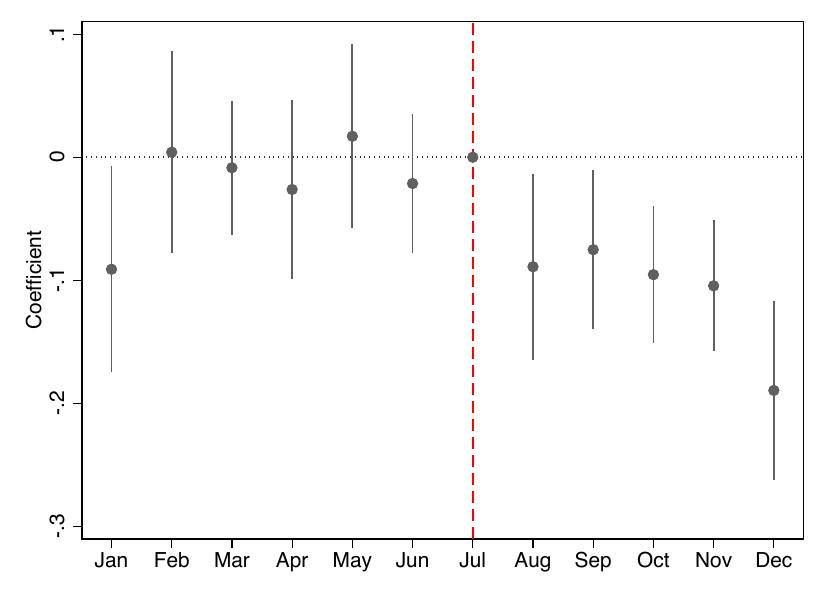}
		\caption{Sunday Afternoon}
		\label{fig:sunday_aft}
	\end{subfigure}

	\label{fig:times}
	\end{center}
\footnotesize{
Notes: The figures present point estimates and 95\% confidence intervals from the event study regressions that include the logarithm of the number of unique monthly visitors. \rev{The sample is the same as that in \Cref{tRegrTimeSubs_walmart_log}.} The dashed vertical line marks the last month before the repeal of the Sunday closing law (July 2019). The coefficient for this month is normalized to zero. Standard errors are clustered at the store level. The p-value from the joint F-test of pre-trends significance of January to June coefficients equals \pretrendPsunmorning{} on \Cref{fig:sunday_morn} and \pretrendPsunaftern{} on \Cref{fig:sunday_aft}.
}
\end{figure}

There is a substantial increase in visits on Sunday mornings following the repeal, indicating that the prior restriction was binding (the first two columns in \Cref{tRegrTimeSubs_walmart_log}). The increase is illustrated in the event study plot in \Cref{fig:sunday_morn}, which shows a sharp upward shift in visits on Sunday morning. The results are robust to alternative functional forms, with the outcome variable in levels or normalized values (see \Cref{tRegrTimeSubs_walmart,tRegrTimeSubs_walmart_log_norm} in the Online Appendix).
\Cref{tRegrTimeSubs_walmart_log} also provides evidence of substitution over time, showing a decline in Sunday afternoon visits. The decline is statistically different from zero, although smaller in magnitude than the Sunday morning increase. It also shows that the policy change did not change the total number of unique visitors.

\begin{table}[ht!]
\renewcommand{\tabcolsep}{3pt}
\begin{center}
\caption{Temporal effects}
\label{tRegrTimeSubs_walmart_log}
\begin{footnotesize}
\begin{tabular}{ l cc cc cc c}
\hline
\hline
 & \multicolumn{2}{c}{Sunday mornings} & \multicolumn{2}{c}{Sunday afternoons} & \multicolumn{2}{c}{Monthly } &  Monthly  \\ & \multicolumn{2}{c}{log visits} & \multicolumn{2}{c}{log visits} &  \multicolumn{2}{c}{log visits} & log visitors\\ & (1) & (2) & (3) & (4) & (5) & (6) & (7)\\ \cmidrule(lr){2-3} \cmidrule(lr){4-5} \cmidrule(lr){6-7} \cmidrule(lr){8-8}
North Dakota $\cdot$ Post&       1.238***&       1.236***&      -0.091***&      -0.093***&       0.019   &       0.017** &       0.003   \\
            &     (0.062)   &     (0.064)   &     (0.018)   &     (0.016)   &     (0.012)   &     (0.007)   &     (0.012)   \\
Log monthly unique visitors &          No   &         Yes   &          No   &         Yes   &          No   &         Yes   &          No   \\
Month FE    &         Yes   &         Yes   &         Yes   &         Yes   &         Yes   &         Yes   &         Yes   \\
Store FE    &         Yes   &         Yes   &         Yes   &         Yes   &         Yes   &         Yes   &         Yes   \\
Dep. var. levels mean in ND, <Aug&        91.5   &        91.5   &       199.7   &       199.7   &      7856.1   &      7856.1   &      3535.4   \\
Stores      &         106   &         106   &         106   &         106   &         106   &         106   &         106   \\
Observations&        1272   &        1272   &        1272   &        1272   &        1272   &        1272   &        1272   \\
\hline

\end{tabular}
\end{footnotesize}
\end{center}
\footnotesize{Notes: The dependent variable is the logarithm of visits and visitors in Walmart stores. The sample includes all Walmart stores in North Dakota and neighboring states, except the two border Walmart stores in Minnesota. Standard errors in parentheses are clustered at the store level. *** Indicates significance at the 1 percent level, ** 5 percent level, * 10 percent level.}
\end{table}

Taken as a whole, these results indicate that when Walmart stores were closed on Sunday mornings, a portion of consumers who would otherwise have shopped during that time shifted to shopping later in the day. Overall, this pattern of substitution reinforces the view that Sunday morning hours were valued, and the law was a binding constraint on consumer behavior.

\subsection{Store Type Substitution}

\begin{figure}[h!]
	\begin{center}
	\caption{Event study of the logarithm of the number of visits to stores on Sunday mornings}

	\begin{subfigure}[b]{0.48\textwidth}
		\includegraphics[width=\textwidth]{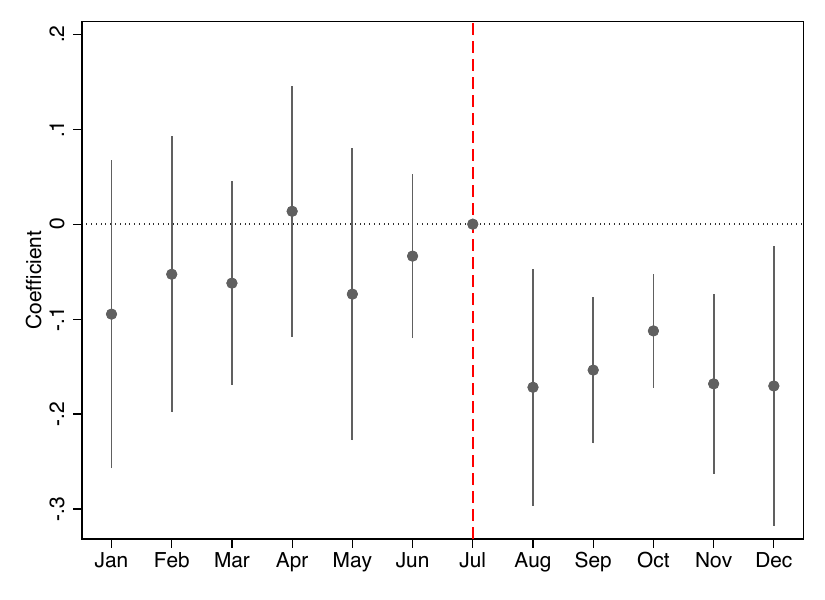}
		\caption{Grocery stores: ND vs. neighbors}
		\label{fig:foodchain_sunday}
	\end{subfigure}
	\hfill
	\begin{subfigure}[b]{0.48\textwidth}
		\includegraphics[width=\textwidth]{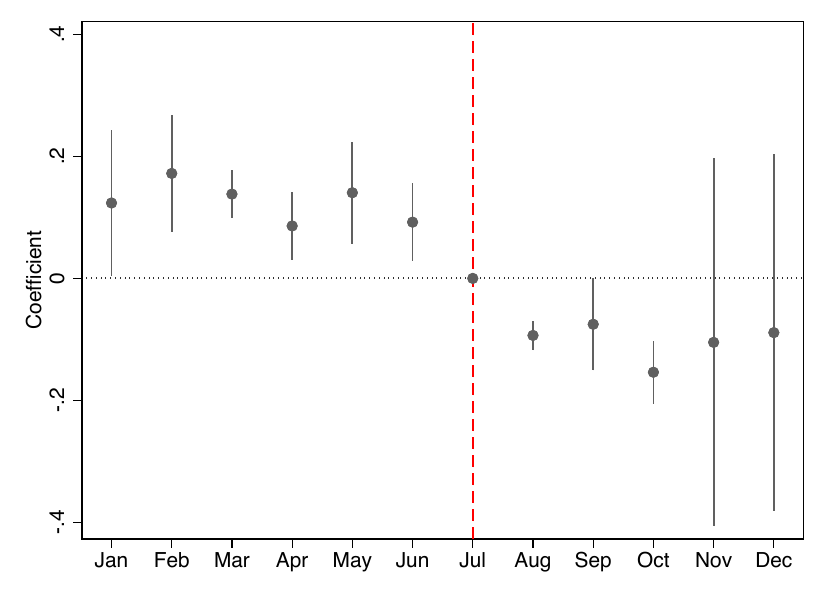}
		\caption{Walmart in MN: bordering ND vs. others}
		\label{fig:border_walmart}
	\end{subfigure}

    \label{fig:heterogeneity}
	\end{center}
\footnotesize{
Notes: The figures present point estimates and 95\% confidence intervals from the event study regressions that include the logarithm of the number of unique monthly visitors.
\rev{\Cref{fig:foodchain_sunday} uses the same sample as columns 1--2 in \Cref{tRegrFood_log}, while \Cref{fig:border_walmart} uses the sample from \Cref{tRegrBorder_log}.}
The dashed vertical line marks the last month before the repeal of the Sunday closing law (July 2019). The coefficient for this month is normalized to zero. Standard errors are clustered at the store level. The p-value from the joint F-test of pre-trends significance of January to June coefficients equals \pretrendPfood{} on \Cref{fig:foodchain_sunday} and \pretrendPborder{} on \Cref{fig:border_walmart}.
}
\end{figure}

Our data allows us to explore two other types of substitution patterns arising from Walmart stores being closed on Sunday mornings. First, we examine whether consumers relocate their visits to different store types in response to the deregulation of opening hours. This can be the case because grocery stores were allowed to be open on Sunday mornings even before the change in policy, unlike Walmart stores. \Cref{tRegrFood_log} (columns 1-2) and \Cref{fig:foodchain_sunday} show that there is a substantial decrease in visits to stores that had a large number of Sunday morning visits before the policy change. We focus on this segment to guarantee that we are looking at food stores that are indeed open on Sunday morning (the fact that they \textit{can} be open does not necessarily imply that they are open). Results suggest that consumers substituted to grocery stores when Walmart stores were closed on Sunday mornings, and are robust to alternative functional forms (see \Cref{tRegrFood,tRegrFood_log_norm} in the Online Appendix). Eye-balling \Cref{fig:foodchain_sunday} reveals no pre-trends prior to August, although the joint F-test of January to June coefficients is statistically significant at ten percent level.

\begin{table}[ht!]
\renewcommand{\tabcolsep}{6pt}
\begin{center}
\caption{Substitution between stores}
\label{tRegrFood_log}
\begin{footnotesize}
\begin{tabular}{ l cc cc cc}
\hline
\hline
 & \multicolumn{4}{c}{Sunday mornings} & \multicolumn{2}{c}{Monthly unique } \\ & \multicolumn{4}{c}{log visits} & \multicolumn{2}{c}{log visitors} \\ & \multicolumn{2}{c}{>90th percentile} & \multicolumn{2}{c}{$\leq$90th percentile} &  >90th & $\leq$90th\\ & (1) & (2) & (3) & (4) & (5) & (6) \\  \cmidrule(lr){2-5} \cmidrule(lr){6-7}
North Dakota $\cdot$ Post&      -0.103** &      -0.112***&      -0.032   &      -0.028   &       0.009   &      -0.004   \\
            &     (0.051)   &     (0.041)   &     (0.035)   &     (0.030)   &     (0.024)   &     (0.026)   \\
Log monthly unique visitors &          No   &         Yes   &          No   &         Yes   &          No   &          No   \\
Month FE    &         Yes   &         Yes   &         Yes   &         Yes   &         Yes   &         Yes   \\
Store FE    &         Yes   &         Yes   &         Yes   &         Yes   &         Yes   &         Yes   \\
Dep. var. levels mean in ND, <Aug&       164.4   &       164.4   &        20.3   &        20.3   &      1721.3   &       361.4   \\
Stores      &          69   &          69   &         660   &         660   &          69   &         660   \\
Observations&         828   &         828   &        7920   &        7920   &         828   &        7920   \\
\hline

\end{tabular}
\end{footnotesize}
\end{center}
\footnotesize{Notes: Outcome variables are the logarithm of the number of visits (columns 1--4) and unique visitors (columns 5--6) in grocery stores. Columns 1--2 restrict the sample to the top 10\% of stores with the largest number of Sunday morning visits before the policy change. Columns 3--4 restrict the sample to the remaining stores.
Standard errors in parentheses are clustered at the store level. *** Indicates significance at the 1 percent level, ** 5 percent level, * 10 percent level.}
\end{table}

\subsection{Cross-Border Substitution}

The second type of substitution we examine is spatial, specifically, cross-border shopping. This mechanism is relevant for the cities of Fargo and Grand Forks, the largest and third-largest cities in North Dakota, respectively. Both cities are located on the state border and form twin-city pairs with Moorhead, Minnesota (adjacent to Fargo), and East Grand Forks, Minnesota (adjacent to Grand Forks). Prior to the repeal of North Dakota’s Sunday closing law, residents of these cities faced restricted shopping opportunities on Sunday mornings. Concurrently, stores across the border in Minnesota remained open. This made cross-border shopping a viable and convenient substitute. We discuss next how the removal of Sunday restrictions in North Dakota affected cross-state substitution. The answer to this question is important, for example, for managers of franchise stores who may consider closing on Sunday while other stores from the same chain, further away, may remain open.

\begin{table}[ht!]
\renewcommand{\tabcolsep}{10pt}
\begin{center}
\caption{Substitution across border}
\label{tRegrBorder_log}
\begin{footnotesize}
\begin{tabular}{ l cc c}
\hline
\hline
 & \multicolumn{2}{c}{Sunday mornings} & Monthly unique \\ & \multicolumn{2}{c}{log visits} & log visitors \\ & (1) & (2) & (3)  \\  \cmidrule(lr){2-3} \cmidrule(lr){4-4}
At North Dakota border $\cdot$ Post&      -0.220** &      -0.210** &      -0.011   \\
            &     (0.107)   &     (0.085)   &     (0.026)   \\
Log monthly unique visitors &          No   &         Yes   &          No   \\
Month FE    &         Yes   &         Yes   &         Yes   \\
Store FE    &         Yes   &         Yes   &         Yes   \\
Dep. var. levels mean at border, <Aug&       310.5   &       310.5   &      2380.8   \\
Stores      &          65   &          65   &          65   \\
Observations&         780   &         780   &         780   \\
\hline

\end{tabular}
\end{footnotesize}
\end{center}
\footnotesize{Notes: Dependent variable is the logarithm of the number of visits and unique visitors in Walmart stores in Minnesota. Standard errors in parentheses are clustered at the store level. *** Indicates significance at the 1 percent level, ** 5 percent level, * 10 percent level.}
\end{table}

\Cref{tRegrBorder_log} (columns 1--2) and \Cref{fig:border_walmart}
show that Sunday morning visits to Walmart stores in Minnesota's border areas declined after the sales restrictions in North Dakota were lifted.
They provide evidence of cross-border shopping substitution arising from the sales restriction. The results are robust to alternative functional forms (see \Cref{tRegrBorder,tRegrBorder_log_norm} in the Online Appendix).
\Cref{fig:border_walmart}  is consistent with a small consumer adjustment already in July. \Cref{fig:border_walmart} also shows large standard errors in November and December, which is explained by the Black Friday and Christmas shopping.

\subsection{Sunday Morning Visits' Attribution} \label{SS:decomosing}

To decompose the estimated increase in Sunday-morning visits, we combine our estimates with the store-count data. The baseline is the estimated per-store increase in Sunday morning visits at North Dakota Walmart stores (\Cref{tRegrTimeSubs_walmart}, Column 2). Temporal substitution is measured by the estimated decline in Sunday afternoon visits at the same stores (\Cref{tRegrTimeSubs_walmart}, Column 4). Cross-border substitution is measured by the estimated decline in visits at Minnesota Walmart stores near the North Dakota border (\Cref{tRegrBorder}, Column 2), which we reallocate proportionally across North Dakota Walmart stores. Grocery store substitution is measured by the estimated decline in Sunday morning visits at North Dakota grocery stores (\Cref{tRegrFood}, Columns 2 and 4), multiplied by the ratio of grocery stores to Walmart stores in the state and summed across the grocery stores assigned to each Walmart. Finally, we measure incremental visits (those that would not have happened if the store was closed) by the total effect on Walmart visits (\Cref{tRegrTimeSubs_walmart}, Column 6) minus the effect on visits to grocery stores or across the border.

Attributing the total effect to sources that we can identify clearly and causally, we find that \decomposeSundayAfternoon{} comes from temporal substitution from Sunday afternoons, \decomposeCrossBorder{} reflects reduced cross-border shopping in Minnesota, and \decomposeGrocery{} captures substitution away from grocery stores that remained open under the old law. The remaining share reflects a combination of additional visits, substitution from other times of the week, and substitution from stores outside our sample, including other chains and non-chain grocery stores. Using the overall change in Walmart visits, we estimate that about \decomposeIncremental{} of the effect is accounted for by options outside our sample, either staying at home or shopping at stores we do not observe. The remaining about \decomposeUnexplained{} reflects substitution from other times of the week.

\section{Welfare Analysis}
\subsection{A Simple Model}

\paragraph{Set-up.}

To quantify the welfare cost of Sunday morning sales restrictions, we build a model where consumers choose stores and shopping time. Each consumer $i$ chooses between four alternatives: visiting a Walmart or a grocery store (which includes supermarkets and convenience stores), either on Sunday morning or at another time. We do not model consumers who don't regularly buy groceries or purchase them from other types of stores. For the model, we use data only from North Dakota. In North Dakota, focusing attention on Walmart and grocery stores is not too restrictive because there are not many other options for grocery shopping.

The utility of consumer $i$ from shopping at a store of type $w=1$ (Walmart) or $w=0$ (grocery store) at time $t=1$ (Sunday morning) or $t=0$ (any other time) is:
\begin{equation}
  U_{iwt} = \alpha D_{iw} + \beta t + \gamma w + \epsilon_{iwt},
\end{equation}
where $D_{iw}$ is the distance from the consumer's home location to the nearest store of type $w$, and $\epsilon_{iwt}$ is an i.i.d. extreme value shock.

Under the standard assumptions of the logit model---namely, that the error terms $\epsilon_{iwt}$ are independent and identically distributed Type I extreme value---the choice shares take the closed form given in equation \eqref{E:shares} (see \cite{train_discrete_2009}, Chapter 3, for derivation). This model yields choice shares across the four alternatives. Specifically, the share of consumer $i$'s visits to store type $w$ at time $t$ is
\begin{equation} \label{E:shares}
s_{iwt}
= \frac{\exp(\alpha D_{iw} + \beta t + \gamma w)}
       {\sum_{w'=0}^1\sum_{t'=0}^1\exp(\alpha D_{iw'} + \beta t' + \gamma w')}.
\end{equation}

\paragraph{Welfare.}

With this model, we can compute the consumer welfare
\begin{equation}
  W = \sum_{i \in I} W_i
    = \sum_{i \in I} \log\sum_{w,t}
      \exp(\alpha D_{iw} + \beta t + \gamma w),
\end{equation}
where $I$ is the set of all consumers.

\paragraph{Counterfactual.}
Our counterfactual is enacting a Sunday morning sales restriction, which removes the option $w=1, t=1$ for all consumers. Let $W^{SM}$ be the welfare in this case. As we remove an option from the choice set, total welfare goes down, but our goal is to compute an interpretable magnitude of this change.

Specifically, we compute the compensating distance, which is a distance $X$ such that moving Walmart stores away by an additional $X$ miles is welfare-equivalent to the inconvenience consumers face due to Sunday sales restrictions.
Let $W^D(X)$ be the welfare in this scenario, where all four options are available to all consumers, but their distance to Walmart is now $D_{i1}+X$ instead of $D_{i1}$. The compensating distance solves the equation

$W^D(X) = W^{SM}$.

Using the model assumptions, we find that the compensating distance is\footnote{
The equation $W_i^D(X) = W_i^{SM}$ simplifies to
\[
\belowdisplayskip=0pt
  \exp\left( \alpha D_{i1} + \alpha X + \beta + \gamma \right)
  +
  \exp\left( \alpha D_{i1} + \alpha X + \gamma \right)
  =
  \exp\left( \alpha D_{i1} + \gamma \right)
\iff
  e^{ \alpha X} \left( 1+e^{\beta} \right)
  =
  1.
\]}
\begin{equation} \label{E:compdist}
X = \frac{\log(1+e^\beta)}{-\alpha}.
\end{equation}

\subsection{Welfare Estimate}

To get a numeric estimate for the compensating distance, we need the values of the three parameters $(\alpha,\beta,\gamma)$. First, using the share of shopping trips that consumer $i$ takes to Walmart, $s_{i1}$, we get\footnote{Derivations of expressions \eqref{E:alphagamma} and \eqref{E:beta} are in \Cref{A:proofs}.}
\begin{equation} \label{E:alphagamma}
\log \frac{s_{i1}}{1-s_{i1}}  = \gamma + \alpha \Delta_i,
\end{equation}
where $\Delta_i = D_{i1}-D_{i0}$ is the differential distance to the closest Walmart vs. the closest grocery store. This allows us to get estimates for the store-type preference parameter $\gamma$ and the distance parameter $\alpha$.

We estimate $\beta$ using the share of Sunday morning visits to store $j$, $s_{j1}$, from
\begin{equation} \label{E:beta}
\log \frac{s_{j1}}{1-s_{j1}}  = \beta.
\end{equation}

A natural concern is whether aggregating to store-level shares introduces bias when consumers have heterogeneous locations. In Appendix B, we show that under our model, the store-level log-odds of Sunday morning visits equals $\beta$ regardless of the distribution of consumer locations, because the time-preference parameter enters additively and cancels out when comparing within-store across time periods. Intuitively, while consumers differ in their distances to stores (which affects their store choice), conditional on visiting a particular store, the relative probability of visiting on Sunday morning versus other times depends only on $\beta$. The aggregation to store-level shares thus preserves identification of the time preference parameter. Finally, note that we observe positive visits in all time periods for all stores in our sample, so corner solutions do not affect our estimation.

The estimate of the slope of the scaled market share of Walmart on the relative distance to Walmart (\Cref{gWalmartShareDistance_ND}) implies $\widehat{\alpha} = -0.02$.
The average logit-transformed share of Sunday morning visits (\Cref{gLogitTransformedSunMorningShareNDPost}) implies $\widehat{\beta} = -3.54$.
Inserting these numbers into \Cref{E:compdist}, we find that the compensating distance is approximately $1.4$ miles.

\subsection{Accounting for Cross-Border Shopping}

The analysis above treated the Sunday closing law as eliminating the option of shopping at Walmart on Sunday mornings. In practice, consumers can instead shop at a Walmart in a neighboring state. For some consumers, this requires a long trip, whereas for those living near the state border, the additional inconvenience is modest.

We incorporate this possibility by modifying the counterfactual: on Sunday mornings, the distance to the relevant Walmart option becomes $D_{i1}+Y_i$ rather than $D_{i1}$, where $Y_i\ge 0$ is the extra travel distance required to reach the closest out-of-state Walmart (relative to the closest in-state Walmart). With this adjustment, the compensating distance for consumer $i$ is
\begin{equation}
  X_i
  =
  \frac{1}{-\alpha}
  \log \left(
  \frac{1+ e^{\beta}}{1+e^{\beta} e^{\alpha Y_i}}
  \right),
\end{equation}
where $\alpha<0$ is the distance coefficient. The impact of the Sunday restriction now depends on location through $Y_i$: the compensating distance $X_i$ increases with $Y_i$.

\begin{figure}[htbp]
  \centering
  \resizebox{0.6\linewidth}{!}{\def\alpha{0.02}
\def\beta{-3.54}
\def\Ymax{220}
\pgfmathsetmacro{\asymptote}{(1/\alpha)*ln(1+exp(\beta))}

\def\matchdistancelist{
  144.39,
  22.27,
  160.21,
  148.33,
  6.89,
  219.38,
  94.75,
  105.86,
  152.45,
  147.96,
  24.21,
  7.93,
  190.04,
  25.41
}

\begin{tikzpicture}
  \begin{axis}[
    width=9cm, height=6.3cm,
    xlabel={Additional distance to out-of-state Walmart (miles)},
    ylabel={Compensating distance (miles)},
    xmin=0, xmax=\Ymax,
    ymin=0, ymax=\asymptote + 0.25,
    axis lines=left,
    tick align=outside,
    grid=major,
    domain=0:\Ymax,
    samples=200,
    clip=false
  ]

    \addplot [black, thick]
      {(1/\alpha)*ln((1+exp(\beta))/(1+exp(\beta)*exp(-\alpha*x)))};

    \addplot [gray, dashed]
      coordinates {(0,\asymptote) (\Ymax,\asymptote)};

    \foreach \y in \matchdistancelist {
      \pgfmathsetmacro{\x}{(1/\alpha)*ln((1+exp(\beta))/(1+exp(\beta)*exp(-\alpha*\y)))}
      \addplot+[
        only marks,
        mark=*,
        mark size=1.8pt,
        black,
        mark options={draw=black, fill=black}
      ] coordinates {(\y,\x)};
    }

  \end{axis}
\end{tikzpicture}}
  \caption{\rev{Compensating distance as a function of additional distance to an out-of-state Walmart. Each dot corresponds to a different Walmart in North Dakota.}}
  \label{F:compdist}
\end{figure}
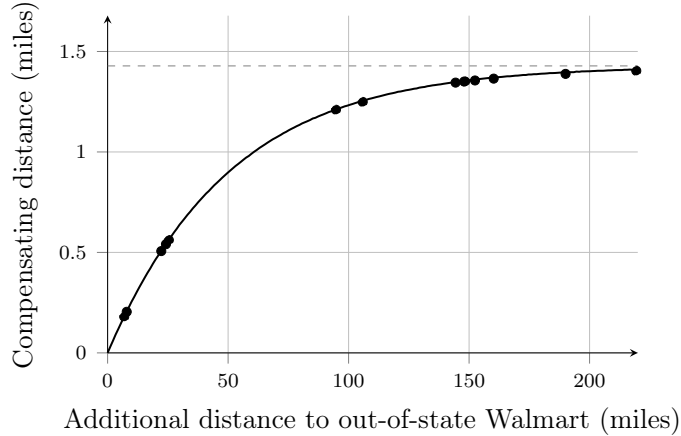

\Cref{F:compdist} illustrates this relationship by plotting $X_i$ against $Y_i$, where each dot corresponds to a North Dakota Walmart and $Y_i$ is computed as the distance from that Walmart to the closest out-of-state Walmart.\footnote{Therefore, the points in the figure can be interpreted as the compensating distances for consumers who live near the corresponding North Dakota store.} For border areas, where the nearest out-of-state Walmart is within 10--20 miles, $X_i$ is substantially smaller than the baseline compensating distance $X$ from the no-substitution counterfactual (which corresponds to $Y_i\to\infty$). For most locations in North Dakota, however, $Y_i$ is large and $X_i$ is close to $X$, because cross-border shopping is impractical.

\section{Concluding Remarks}

We studied the impact of store opening hour deregulation on consumer behavior and welfare using the 2019 repeal of North Dakota’s Sunday closing law as a natural experiment. We showed that lifting the restriction led to a substantial increase in Sunday morning visits. We documented intertemporal substitution, substitution across store types, and cross-border substitution.

Our evidence shows that Sunday-morning trading restrictions impose real costs on consumers, even where alternative shopping options exist.
To quantify the welfare implications of the regulation, we presented a framework to model consumers’ preferences over the three documented dimensions: time, store type, and distance. The welfare analysis suggests that deregulating Sunday morning store hours is equivalent to a decrease in the travel distance to Walmart stores by 1.4 miles for each consumer.

It is worth emphasizing that this welfare analysis captures only the consumer side of the policy’s effects. We do not analyze potential changes in firm behavior, such as pricing, product variety, investment, or labor-market outcomes. Our findings measure only one side of the policy’s impact, while all the above factors are important for the overall welfare.

Our findings have important implications for both policymakers and store managers.
Policymakers face a trade-off: restrictions may help maintain a shared day of rest and can benefit small retailers or workers seeking more convenient schedules, but they also reduce flexibility for time-constrained households and may shift spending to other jurisdictions.
Our paper provides clear evidence regarding how consumers are affected by such regulations. Likewise, managers have to decide the days and time slots in which they will be open to the public. They also need to decide whether to have a uniform chain-wide policy or to allow each store to have a separate policy. Understanding how many visits they will lose if the store is closed on Sunday, and if those visits are lost to competitors or just relocated to other days of the week, is a crucial input into a manager's store hour policy.

\clearpage
\appendix

\counterwithin{table}{section}
\counterwithin{figure}{section}
\setcounter{page}{1}\newpage
\renewcommand{\thepage}{A\arabic{page}}

\part*{Appendix}

\section{Additional Figures and Tables}\label{A:AdditionalTablesFigures}

\begin{figure}[ht]
	\centering
	\includegraphics[width=0.8\textwidth]{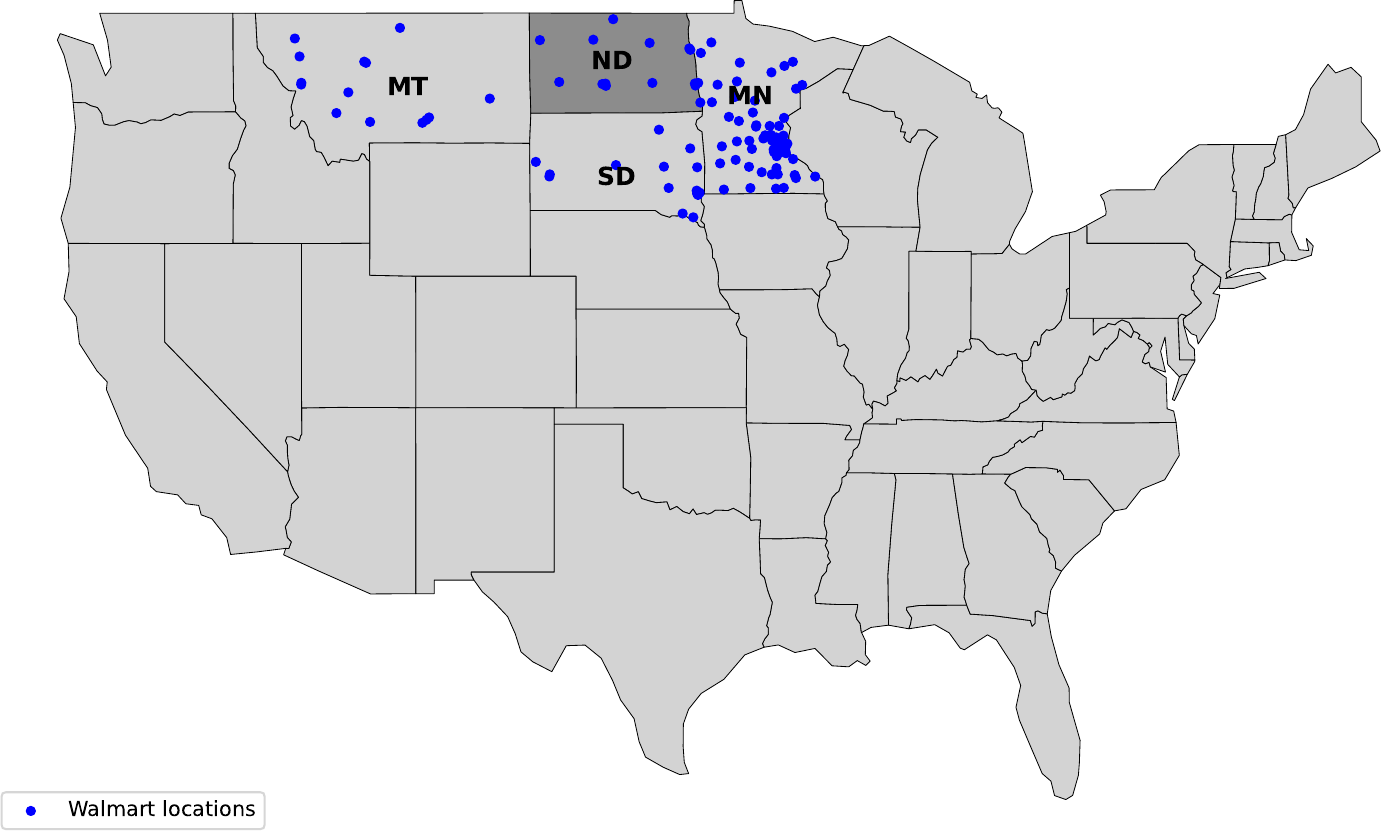}
	\caption{North Dakota and neighboring comparison states used in the analysis. Blue markers show Walmart superstore locations.}
	\label{fig:map_all_states}
\end{figure}

\begin{table}[ht!]
\renewcommand{\tabcolsep}{3pt}
\begin{center}
\caption{Temporal effects. Visits and visitors in Walmart stores}
\label{tRegrTimeSubs_walmart}
\begin{footnotesize}
\begin{tabular}{ l cc cc cc c}
\hline
\hline
 & \multicolumn{2}{c}{Sunday mornings} & \multicolumn{2}{c}{Sunday afternoons} & \multicolumn{2}{c}{Monthly } &  Monthly  \\ & \multicolumn{2}{c}{ visits} & \multicolumn{2}{c}{ visits} &  \multicolumn{2}{c}{ visits} &  visitors\\ & (1) & (2) & (3) & (4) & (5) & (6) & (7)\\ \cmidrule(lr){2-3} \cmidrule(lr){4-5} \cmidrule(lr){6-7} \cmidrule(lr){8-8}
North Dakota $\cdot$ Post&     207.733***&     204.108***&     -14.623***&     -15.824***&     220.560   &     152.170** &      42.393   \\
            &    (34.976)   &    (31.786)   &     (4.157)   &     (3.821)   &   (152.967)   &    (74.904)   &    (71.864)   \\
Monthly unique visitors &          No   &         Yes   &          No   &         Yes   &          No   &         Yes   &          No   \\
Month FE    &         Yes   &         Yes   &         Yes   &         Yes   &         Yes   &         Yes   &         Yes   \\
Store FE    &         Yes   &         Yes   &         Yes   &         Yes   &         Yes   &         Yes   &         Yes   \\
Dep. var. levels mean in ND, <Aug&        91.5   &        91.5   &       199.7   &       199.7   &      7856.1   &      7856.1   &      3535.4   \\
Stores      &         106   &         106   &         106   &         106   &         106   &         106   &         106   \\
Observations&        1272   &        1272   &        1272   &        1272   &        1272   &        1272   &        1272   \\
\hline

\end{tabular}
\end{footnotesize}
\end{center}
\footnotesize{Notes: Standard errors in parentheses are clustered at the store level. *** Indicates significance at the 1 percent level, ** 5 percent level, * 10 percent level.}
\end{table}

\begin{table}[ht!]
\renewcommand{\tabcolsep}{3pt}
\begin{center}
\caption{Temporal effects. Logarithm of normalized visits in Walmart stores}
\label{tRegrTimeSubs_walmart_log_norm}
\begin{footnotesize}
\begin{tabular}{ l cc cc cc}
\hline
\hline
 & \multicolumn{2}{c}{Sunday mornings} & \multicolumn{2}{c}{Sunday afternoons} & \multicolumn{2}{c}{Monthly } \\ & \multicolumn{2}{c}{log visits} & \multicolumn{2}{c}{log visits} &  \multicolumn{2}{c}{log visits} \\ & (1) & (2) & (3) & (4) & (5) & (6) \\ \cmidrule(lr){2-3} \cmidrule(lr){4-5} \cmidrule(lr){6-7}
North Dakota $\cdot$ Post&       1.249***&       1.246***&      -0.091***&      -0.094***&       0.019   &       0.017** \\
            &     (0.063)   &     (0.064)   &     (0.018)   &     (0.016)   &     (0.012)   &     (0.008)   \\
Log monthly unique visitors &          No   &         Yes   &          No   &         Yes   &          No   &         Yes   \\
Month FE    &         Yes   &         Yes   &         Yes   &         Yes   &         Yes   &         Yes   \\
Store FE    &         Yes   &         Yes   &         Yes   &         Yes   &         Yes   &         Yes   \\
Dep. var. levels mean in ND, <Aug&      1292.4   &      1292.4   &      2826.9   &      2826.9   &    110998.1   &    110998.1   \\
Stores      &         106   &         106   &         106   &         106   &         106   &         106   \\
Observations&        1272   &        1272   &        1272   &        1272   &        1272   &        1272   \\
\hline

\end{tabular}
\end{footnotesize}
\end{center}
\footnotesize{Notes: Standard errors in parentheses are clustered at the store level. *** Indicates significance at the 1 percent level, ** 5 percent level, * 10 percent level.}
\end{table}

\begin{table}[ht!]
\renewcommand{\tabcolsep}{6pt}
\begin{center}
\caption{Substitution between stores. Visits and visitors in grocery stores}
\label{tRegrFood}
\begin{footnotesize}
\begin{tabular}{ l cc cc cc}
\hline
\hline
 & \multicolumn{4}{c}{Sunday mornings} & \multicolumn{2}{c}{Monthly unique } \\ & \multicolumn{4}{c}{ visits} & \multicolumn{2}{c}{ visitors} \\ & \multicolumn{2}{c}{>90th percentile} & \multicolumn{2}{c}{$\leq$90th percentile} &  >90th & $\leq$90th\\ & (1) & (2) & (3) & (4) & (5) & (6) \\  \cmidrule(lr){2-5} \cmidrule(lr){6-7}
North Dakota $\cdot$ Post&     -26.636***&     -26.970***&      -3.412***&      -2.015***&       2.192   &     -15.943*  \\
            &     (8.195)   &     (6.670)   &     (1.037)   &     (0.696)   &    (41.348)   &     (8.146)   \\
Monthly unique visitors &          No   &         Yes   &          No   &         Yes   &          No   &          No   \\
Month FE    &         Yes   &         Yes   &         Yes   &         Yes   &         Yes   &         Yes   \\
Store FE    &         Yes   &         Yes   &         Yes   &         Yes   &         Yes   &         Yes   \\
Dep. var. levels mean in ND, <Aug&       164.4   &       164.4   &        20.3   &        20.3   &      1721.3   &       361.4   \\
Stores      &          69   &          69   &         660   &         660   &          69   &         660   \\
Observations&         828   &         828   &        7920   &        7920   &         828   &        7920   \\
\hline

\end{tabular}
\end{footnotesize}
\end{center}
\footnotesize{Notes: Columns 1--2 restrict the sample to the top 10\% of stores with the largest number of Sunday morning visits before the policy change. Columns 3--4 restrict the sample to the remaining stores.
Standard errors in parentheses are clustered at the store level. *** Indicates significance at the 1 percent level, ** 5 percent level, * 10 percent level.}
\end{table}

\begin{table}[ht!]
\renewcommand{\tabcolsep}{6pt}
\begin{center}
\caption{Substitution between stores. Logarithm of normalized visits in grocery stores}
\label{tRegrFood_log_norm}
\begin{footnotesize}
\begin{tabular}{ l cc cc}
\hline
\hline
 & \multicolumn{4}{c}{Sunday mornings} \\ & \multicolumn{4}{c}{log visits} \\ & \multicolumn{2}{c}{>90th percentile} & \multicolumn{2}{c}{$\leq$90th percentile} \\ & (1) & (2) & (3) & (4)  \\  \cmidrule(lr){2-5}
North Dakota $\cdot$ Post&      -0.104** &      -0.113***&      -0.006   &      -0.001   \\
            &     (0.052)   &     (0.042)   &     (0.048)   &     (0.040)   \\
Log monthly unique visitors &          No   &         Yes   &          No   &         Yes   \\
Month FE    &         Yes   &         Yes   &         Yes   &         Yes   \\
Store FE    &         Yes   &         Yes   &         Yes   &         Yes   \\
Dep. var. levels mean in ND, <Aug&      2325.6   &      2325.6   &       287.3   &       287.3   \\
Stores      &          69   &          69   &         660   &         660   \\
Observations&         828   &         828   &        7920   &        7920   \\
\hline

\end{tabular}
\end{footnotesize}
\end{center}
\footnotesize{Notes: Columns 1--2 restrict the sample to the top 10\% of stores with the largest number of Sunday morning visits before the policy change. Columns 3--4 restrict the sample to the remaining stores.
Standard errors in parentheses are clustered at the store level. *** Indicates significance at the 1 percent level, ** 5 percent level, * 10 percent level.}
\end{table}

\begin{table}[ht!]
\renewcommand{\tabcolsep}{20pt}
\begin{center}
\caption{Substitution across border. Visits and visitors in Walmart stores in Minnesota}
\label{tRegrBorder}
\begin{footnotesize}
\begin{tabular}{ l cc c}
\hline
\hline
 & \multicolumn{2}{c}{Sunday mornings} & Monthly unique \\ & \multicolumn{2}{c}{ visits} &  visitors \\ & (1) & (2) & (3)  \\  \cmidrule(lr){2-3} \cmidrule(lr){4-4}
At North Dakota border $\cdot$ Post&    -106.706***&     -88.482** &    -203.466***\\
            &    (36.365)   &    (37.491)   &    (37.303)   \\
Monthly unique visitors &          No   &         Yes   &          No   \\
Month FE    &         Yes   &         Yes   &         Yes   \\
Store FE    &         Yes   &         Yes   &         Yes   \\
Dep. var. levels mean at border, <Aug&       310.5   &       310.5   &      2380.8   \\
Stores      &          65   &          65   &          65   \\
Observations&         780   &         780   &         780   \\
\hline

\end{tabular}
\end{footnotesize}
\end{center}
\footnotesize{Notes: Standard errors in parentheses are clustered at the store level. *** Indicates significance at the 1 percent level, ** 5 percent level, * 10 percent level.}
\end{table}

\begin{table}[ht!]
\renewcommand{\tabcolsep}{20pt}
\begin{center}
\caption{Substitution across border. Logarithm of normalized visits in Walmart stores in Minnesota}
\label{tRegrBorder_log_norm}
\begin{footnotesize}
\begin{tabular}{ l cc}
\hline
\hline
 & \multicolumn{2}{c}{Sunday mornings}  \\ & \multicolumn{2}{c}{log visits}  \\ & (1) & (2)  \\  \cmidrule(lr){2-3}
At North Dakota border $\cdot$ Post&      -0.222** &      -0.212** \\
            &     (0.108)   &     (0.085)   \\
Log monthly unique visitors &          No   &         Yes   \\
Month FE    &         Yes   &         Yes   \\
Store FE    &         Yes   &         Yes   \\
Dep. var. levels mean at border, <Aug&      4407.9   &      4407.9   \\
Stores      &          65   &          65   \\
Observations&         780   &         780   \\
\hline

\end{tabular}
\end{footnotesize}
\end{center}
\footnotesize{Notes: Standard errors in parentheses are clustered at the store level. *** Indicates significance at the 1 percent level, ** 5 percent level, * 10 percent level.}
\end{table}

\begin{figure}[ht]
	\begin{center}
	\includegraphics[width=0.6\textwidth]{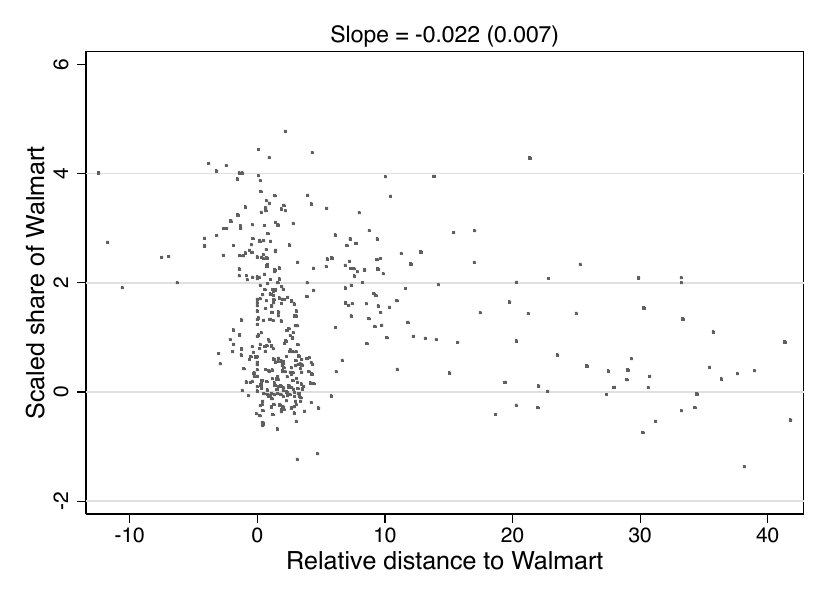}
	\caption{Relative (to grocery store) distance to Walmart vs logit-transformed market share of Walmart in North Dakota after the change in regulation}
	\label{gWalmartShareDistance_ND}
	\end{center}
\footnotesize{
Notes: The figure presents a scatter plot, where an observation is a census block group in North Dakota post deregulation. The x-axis measures the distance to the closest Walmart minus the distance to the closest grocery store. The y-axis measures the logit-transformed market share of Walmart $log(s_{iW}/(1-s_{iW}))$, where $s_{iW}$ is the share of Walmart from all grocery and Walmart visits from census block group $i$. To reduce the noise, we exclude from the sample census block groups in the lowest decile in terms of the average number of devices.
}
\end{figure}

\begin{figure}[ht]
	\begin{center}
	\includegraphics[width=0.6\textwidth]{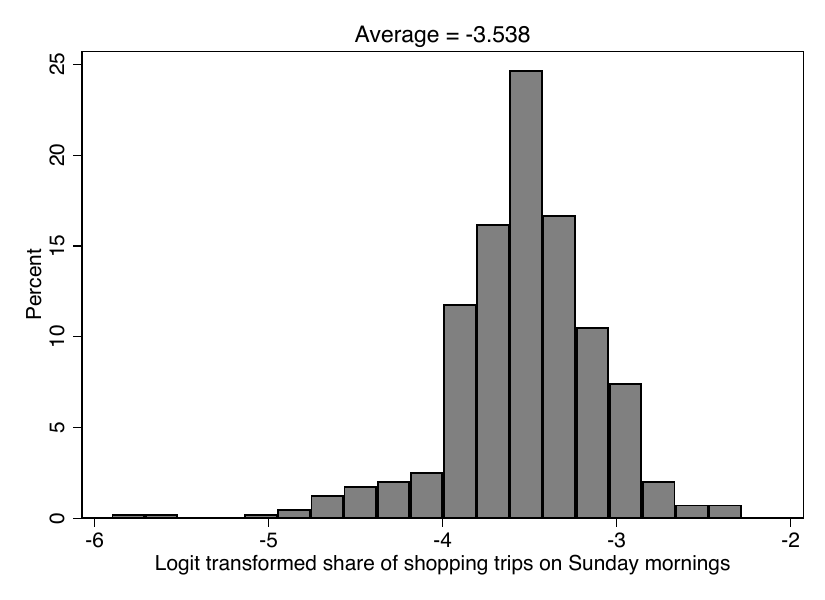}
	\caption{The distribution of logit transformed share of Sunday morning store visits in North Dakota after the change in regulation}
	\label{gLogitTransformedSunMorningShareNDPost}
	\end{center}
\footnotesize{
Notes: The figure presents the histogram of the logit transformed share of Sunday morning store visits  $\log(s_{jt}/(1-s_{jt}))$, where $s_{jt}$ is the share of visits to store $j$ in month $t$ on Sunday mornings. The sample includes both Walmart stores and grocery stores in weeks post-deregulation.
}
\end{figure}

\clearpage
\section{Derivations} \label{A:proofs}

Using equation \eqref{E:shares}, we can write the share of consumer $i$'s visit to Walmart as
\[
s_{i1}
=
s_{i10}
+
s_{i11}
= \frac{
\exp(\alpha D_{i1} + \gamma) (1+e^{\beta})
}
       {\sum_{w'=0}^1\sum_{t'=0}^1\exp(\alpha D_{iw'} + \beta t' + \gamma w')}.
\]
The share of the visits to stores other than Walmart is
\[
s_{i0}=1-s_{i1}=s_{i00}+s_{i01}
= \frac{
\exp(\alpha D_{i0}) (1+e^{\beta})
}
       {\sum_{w'=0}^1\sum_{t'=0}^1\exp(\alpha D_{iw'} + \beta t' + \gamma w')}.
\]
Taking the difference of the logarithms of these two shares, we get
\[
\log \frac{s_{i1}}{1-s_{i1}}
= \log s_{i1} - \log s_{i0}
= \gamma + \alpha \Delta_i,
\]
where $\Delta_i = D_{i1} - D_{i0}$. This is \eqref{E:alphagamma}, which we can directly take to the data.

Now, if we fix a store (either Walmart or grocery store) and take a consumer $i$ for whom it is the closest store of the respective type, then the share of this consumer's visits that occur on Sunday is
\[
s_{iw1}
= \frac{\exp(\alpha D_{iw} + \beta + \gamma w)}
       {\sum_{w'=0}^1\sum_{t'=0}^1\exp(\alpha D_{iw'} + \beta t' + \gamma w')}.
\]
And the share of visits that occur at all other times is
\[
s_{iw0}
= \frac{\exp(\alpha D_{iw} + \gamma w)}
       {\sum_{w'=0}^1\sum_{t'=0}^1\exp(\alpha D_{iw'} + \beta t' + \gamma w')}.
\]
Therefore, the log difference is
\[
\log \frac{s_{iw1}}{s_{iw0}}
=
\log s_{iw1} - \log s_{iw0}
= (\alpha D_{iw} + \beta + \gamma w)-(\alpha D_{iw} + \gamma w)=\beta.
\]

If we could observe shopping times of each consumer separately, this would be enough to estimate $\beta$. However, in the data, we only observe this information at the store level, i.e., in store $j$, the share of visits that occurred on Sunday morning vs. other times.

To find this aggregate and see that it does not complicate the analysis, let us introduce some notation. First, let us fix a store $j$ and let its type be $w_j \in \{0,1\}$. Let the set of consumers for whom it is the closest store of type $w_j$ be denoted by $I_j$. For each such consumer, let $n_i$ be the number of shopping trips by consumer $i$.

With this notation, we can compute the total number of shopping trips to the store $j$ that occur at times other than Sunday morning as
\begin{align*}
N_{j0}
&= \sum_{i \in I_j} n_i s_{iw_j0}
= \sum_{i \in I_j} n_i \frac{\exp(\alpha D_{iw_j} + \gamma w_j)}
       {\sum_{w'=0}^1\sum_{t'=0}^1\exp(\alpha D_{iw'} + \beta t' + \gamma w')}.
\end{align*}
Similarly, we can compute the total number of shopping trips to store $j$ on Sunday morning as
\begin{align*}
N_{j1}
&= \sum_{i \in I_j} n_i s_{iw_j1}
= \sum_{i \in I_j} n_i \frac{\exp(\alpha D_{iw_j} + \beta + \gamma w_j)}
       {\sum_{w'=0}^1\sum_{t'=0}^1\exp(\alpha D_{iw'} + \beta t' + \gamma w')}
\\
&=
 \exp(\beta)
\sum_{i \in I_j} n_i \frac{\exp(\alpha D_{iw_j}+\gamma w_j)}
       {\sum_{w'=0}^1\sum_{t'=0}^1\exp(\alpha D_{iw'} + \beta t' + \gamma w')}
= e^{\beta} N_{j0}.
\end{align*}
Therefore, the respective shares are
\[
s_{j0}
= \frac{N_{j0}}{N_{j0}+N_{j1}}
= \frac{1}{1+e^{\beta}}
,
\;\;
s_{j1}
= \frac{N_{j1}}{N_{j0}+N_{j1}}
= \frac{e^{\beta}}{1+e^{\beta}}.
\]
Computing the log difference gives us
\[
\log \frac{s_{j1}}{1-s_{j1}}
= \log s_{j1} - \log s_{j0}
= [\beta - \log(1+e^{\beta})] - [0 - \log(1+e^{\beta})]
= \beta.
\]
This is \eqref{E:beta}.

\end{document}